\documentclass[iop]{emulateapj}

\usepackage{graphicx}	
\usepackage{amsmath}	

\usepackage{amssymb}	
\usepackage{bm}
\usepackage{booktabs}
\usepackage{cleveref}
\usepackage{natbib}
\usepackage{threeparttable}

\newcommand{\secpoint}{\mbox{$''\mskip-7.6mu.\,$}}
\newcommand{\angstrom}{\mbox{\normalfont\AA}}

\begin{document}

\title{The Detection of Ionized Carbon Emission at $\lowercase{z}\sim 8$\altaffilmark{1}}

\author{
Michael W. Topping,\altaffilmark{2}
Alice E. Shapley,\altaffilmark{3}
Daniel P. Stark,\altaffilmark{2}
Ryan Endsley,\altaffilmark{2}
Brant Robertson,\altaffilmark{4}
Jenny E. Greene,\altaffilmark{5}
Steven R. Furlanetto,\altaffilmark{3}
Mengtao Tang\altaffilmark{6}
}

\altaffiltext{1}{Based on data obtained at the W.M. Keck Observatory, which is operated as a scientific partnership among the California Institute of Technology, the University of California,  and the National Aeronautics and Space Administration, and was made possible by the generous financial support of the W.M. Keck Foundation.}
\altaffiltext{2}{Steward Observatory, University of Arizona, 933 N Cherry Ave, Tucson, AZ 85721, USA}
\altaffiltext{3}{Department of Physics and Astronomy, University of California, Los Angeles, 430 Portola Plaza, Los Angeles, CA 90095, USA}
\altaffiltext{4}{Department of Astronomy and Astrophysics, University of California, Santa Cruz, 1156 High Street, Santa Cruz, CA 95064, USA}
\altaffiltext{5}{Department of Astrophysical Sciences, Princeton University, 4 Ivy Lane, Princeton, NJ, 08544, USA}
\altaffiltext{6}{Department of Physics and Astronomy, University College London, Gower Street, London WC1E 6BT, UK}
\email{michaeltopping@arizona.edu}

\shortauthors{Topping et al.}

\shorttitle{Ionized Carbon Emission at $\lowercase{z}\sim 8$}


\begin{abstract}
We present deep Keck/MOSFIRE $H$-band spectroscopic observations covering the [CIII],CIII]$\lambda\lambda1907,1909$ doublet for three $z\sim8$ galaxy candidates in the AEGIS field.  Along with non-detections in two galaxies, we obtain one of the highest-redshift detections to-date of [CIII]$\lambda 1907$ for the galaxy AEGIS-33376, finding $z_{\rm spec}=7.945\pm0.001$. 
We measure a  [CIII]$\lambda$1907 flux of $2.24\pm0.71\times10^{-18} \mbox{ erg}\mbox{ s}^{-1} \mbox{ cm}^{-2}$, 
corresponding to a rest-frame equivalent width of $20.3\pm6.5\ \mbox{\AA}$ for the single line.  
Given the not very constraining upper limit for CIII]$\lambda 1909$ based on strong sky-line contamination, we assume a [CIII]$\lambda$1907/CIII]$\lambda 1909$ doublet ratio of 1.5 and infer a total [CIII],CIII]$\lambda\lambda1907,1909$ equivalent width of $33.7\pm 10.8\ \mbox{\AA}$. 
We repeat the same reductions and analysis on multiple subsets of our raw data divided on the basis of time and observing conditions, verifying that the [CIII]$\lambda 1907$ emission is present for AEGIS-33376 throughout our observations. We also confirm that the significance of the [CIII]$\lambda 1907$ detection in different subsets of our data tracks that of brighter emission features detected on the same multi-slit mask. These multiple  tests suggest that the observed emission line is real and associated with the $z\sim 8$ target.  The strong observed [CIII],CIII]$\lambda\lambda1907,1909$ in AEGIS-33376 likely indicates ISM conditions of low metallicity, high ionization parameter, and a hard ionizing spectrum, although AGN contributions are possible.  This single detection represents a sizable increase in the current sample [CIII],CIII]$\lambda\lambda1907,1909$ detections at $z>7$, while {\it JWST} will provide the first statistical samples of such measurements at these redshifts.


\end{abstract}


%
%
\section{Introduction} 
\label{sec:intro}
Understanding the properties of the earliest 
galaxies and their role in the reionization of the universe comprises one of the current frontiers for extragalactic astronomy.  Because of the apparent faintness of galaxies from the reionization epoch, and the associated challenge of observing such objects, much of our understanding of the galaxy population in the $z>7$ universe to date, including star formation rates, star formation histories, and stellar masses, is based on photometric measurements \citep[e.g.,][]{Labbe2013, Bouwens2014, Bouwens2015, Finkelstein2015, Stark2016}. 
Such broadband photometric measurements can also be used to estimate properties including the strengths of prominent rest-frame optical emission lines \citep{Ono2012, Smit2014, Smit2015, Roberts-Borsani2016, deBarros2019, Endsley2021a}, but these still only represent indirect inferences.
Indeed, while many features of the galaxy population during the reionization epoch have been described with photometry, such methods are not ideal when trying to establish and understand the detailed internal physical processes that govern galaxy evolution.

Obtaining the spectra of galaxies at the highest redshifts provides the most effective tool for understanding detailed galaxy properties.  At a very basic level, spectroscopic redshift measurements yield distances that enable the precise determination of intrinsic luminosities and rest-frame colors. 
In addition to redshift measurements, spectroscopic observations provide windows into the properties of massive stars and the interstellar medium during crucial evolutionary periods in the history of the universe, such as cosmic reionization.  During this early epoch, strong rest-optical emission lines are redshifted to wavelengths that are inaccessible from the ground. Accordingly, measurements in the rest-UV and submillimeter comprise the best options for inferring galaxy properties through spectroscopic means. One key feature is [CIII],CIII]$\lambda\lambda1907,1909$, which is often the second strongest emission feature in the rest-UV, after H~I Ly$\alpha$ \citep{Shapley2003}. The CIII] doublet --  in combination with C~IV$\lambda\lambda1548,1550$, He~II$\lambda1640$, [OIII]$\lambda\lambda1661,1666$, and SiIII]$\lambda\lambda1882,1892$ -- also provides an estimate of the ionizing radiation field and gas-phase metallicity within the ISM \citep[see e.g.,][]{Byler2018}.  However, observing these faint emission lines is difficult, and only a small number of detections have been achieved to date \citep{Stark2015-c3, Stark2015-c4, Mainali2017, Stark2017, Laporte2017, Hutchison2019, Jiang2021}. Thus far, measurements of these rest-UV features at high redshifts have implied ISM properties consistent with low metallicities, high ionization parameters, and the presence of a hard ionizing spectrum.

The next generation of ground-based ELTs will enable us to probe to fainter emission-line fluxes, covering a larger range of galaxy properties.  In addition, the \textit{James Webb Space Telescope} (JWST) will allow observation of strong rest-optical emission lines that are currently inaccessible from the ground.  With these facilities, observations of faint emission lines during the reionization era will become standard, and the detection of many emission features for a statistical sample of typical galaxies during this epoch will come within reach.  In the meantime, it is still possible,
although extremely challenging, to push the current generation of 10-meter-class ground-based telescopes and sensitive near-infrared spectrographs and pursue the rest-UV emission features of $z>7$ galaxies. Here we present deep near-infrared spectroscopic measurements of three  $z>7$ galaxy candidates with $\rm M_{\rm UV}$$\sim$$-21$ using the MOSFIRE spectrograph \citep{mclean2012} on the Keck~I telescope, including a detection of [CIII]$\lambda1907$ emission at $z=7.945$.

This paper is organized as follows.  Section 2 outlines our observations and reductions.  Section 3 presents our main results, and Section 4 provides a brief discussion and summary.  Throughout this paper, we adopt cosmological parameters of $H_0=70\mbox{ km}\mbox{ s}^{-1}\mbox{ Mpc}^{-1}$, $\Omega_M = 0.3$, and $\Omega_{\Lambda}=0.7$.

%
%
\section{Data and Methods}
\label{sec:data}

\begin{figure*}
    \centering
    \includegraphics[width=1.0\linewidth]{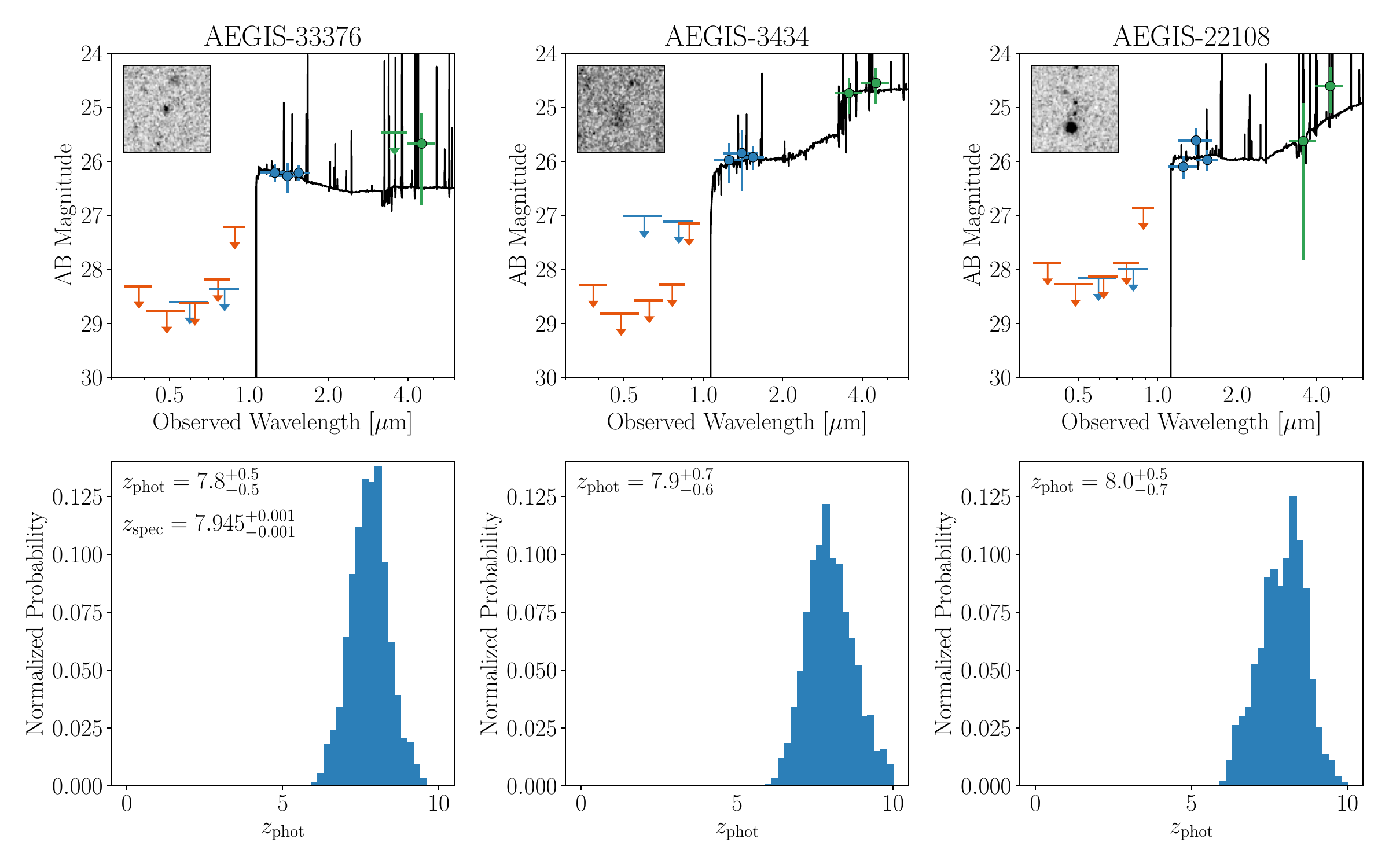}
    \caption{\textit{Top:} Best-fit SEDs calculated using BEAGLE \citep{Chevallard2016} for the three objects in our sample. Observed photometric measurements and $2\sigma$ upper limits are shown in blue for \textit{HST} ACS/WFC3 \citep{Grogin2011, Koekemoer2011, Skelton2014}, green for \textit{Spitzer} IRAC bands \citep{Ashby2015}, and red for ground-based CFHT data \citep{Gwyn2012}.  Inset into each panel are cutouts $5^{\prime \prime}$ on a side centered on the position of each object from the \textit{HST} F160W image. \textit{Bottom:} Corresponding photometric redshift probability distribution functions.}
    \label{fig:SEDs}
\end{figure*}

\subsection{Observations and Reductions}
We obtained near-infrared spectroscopic observations in the AEGIS field using  the Keck/MOSFIRE spectrograph \citep{mclean2012}. MOSFIRE slitmasks were filled with a variety of targets benefiting from long exposure times. These primarily included star-forming galaxies at $z\sim1.5$ for which we were attempting deep observations of [NII]$\lambda6584$/H$\alpha$ to achieve a high detection fraction of the faint [NII]$\lambda6584$ feature. We also targeted galaxy candidates at $z>7$ drawn from the catalogs of \citet{Bouwens2015}. We constructed broadband SEDs for each of our $z>7$ targets, utilizing multi-wavelength photometric data that were assembled for the AEGIS field \citep{Stefanon2017}. The best-fit SEDs were derived using the BEAGLE SED fitting code, with which we also recalculated the photometric redshift probability distribution functions \citep{Chevallard2016}. BEAGLE utilizes the latest \citet{Bruzual2003} stellar population models and incorporates self-consistent photoionization models \citep{Gutkin2016} using Cloudy \citep{Ferland2017}.
The three most robust $z>7$ candidates were AEGIS-33376, AEGIS-3434, and AEGIS-22108 (Figure~\ref{fig:SEDs}), where the identification numbers are taken from the 3D-HST v4.1 catalog \citep{Skelton2014}. 

MOSFIRE data were obtained through two different slitmasks in the AEGIS field, one (ae1\_w1) containing 33376, and the other (ae1\_w2)  containing 3434 and 22108. 
The ae1\_w1 mask was observed on 2020 March 11 and 12, and 2021 May 1 and 2 (UT), while the ae1\_w2 mask was observed on 2021 April 19 and 2021 May 1 (UT). The AEGIS masks were observed in the $H$ band with 0\secpoint7 slits, yielding a spectral resolution of $\sim3650$. We integrated for a total of 806 and 240 minutes, respectively, on the ae1\_w1 and ae1\_w2  masks. Individual exposures were 120 s, and followed an ABA'B' dither pattern characterized by, respectively, 1\secpoint5 and 1\secpoint2 outer and inner nods. Conditions were clear, and the median seeing for both masks was 0\secpoint6.

The data were reduced using a custom IDL pipeline that yields two-dimensional rectified, flat-fielded, sky-subtracted, cosmic-ray-cleaned, wavelength-calibrated and flux-calibrated science and error spectra for each slitlet \citep{kriek2015}. One-dimensional science and error spectra were optimally extracted and corrected for slit losses using {\it Hubble Space Telescope} ({\it HST}) F160W images and the median seeing for each mask.

\subsection{Spectrum extraction}

For objects on the masks with visible emission lines, we obtained 1D spectra from the reduced images by performing an optimal extraction defined by fitting a Gaussian to the spatial profile of the emission line \citep{Horne1986}. If no emission lines were visible, we blindly extracted a spectrum based on the expected position in the slit with a width defined by the average seeing measured from a brighter (AB mag=19--20) star on the mask.  Limiting line fluxes for non-detections were estimated by integrating the square of the error spectrum over a typical line width centered on each pixel, and taking the square root of the resulting value.  
For objects with no spectroscopic redshift, the line flux upper limit is not clearly defined, since the noise is dominated by the sky spectrum and a very strong function of wavelength (i.e., redshift). To reflect this dependence and report {\it typical} line flux uncertainties, we estimated the 5$\sigma$ line-flux limit as a function of wavelength over the range of possible observed wavelengths of CIII] based on the photometric redshift uncertainty. We then report the median 5$\sigma$ line-flux limit over this wavelength range as the CIII] flux error for non-detections with only photometric redshifts. 
Upper limits for the combined [CIII],CIII]$\lambda\lambda1907,1909$ doublet were measured using a similar procedure, however we performed the integration over two wavelength windows with a fixed separation to match the doublet members.

%
%
\section{Results}
\label{sec:results}

We visually inspect the 2D $H$-band spectra of all three $z>7$ targets to find emission lines. These spectra cover an observed wavelength range of $\rm 1.47\mu m-1.80\mu m$, $\rm 1.45\mu m-1.76\mu m$, and $\rm 1.51\mu m-1.81\mu m$ for AEGIS-33376, AEGIS-3434, and AEGIS-22108 respectively.  We identify one emission line in the spectrum of AEGIS-33376 at $\lambda=17055\angstrom$ at the expected spatial position of the galaxy, and do not find any detections in the spectrum of either AEGIS-3434 or AEGIS-22108. The detected emission line in AEGIS-33376 is coincident with a sky line. However, this sky line is one of the weakest in the $H$ band.  Despite the increased noise due to sky contamination, we measure a flux of $2.24\pm0.71\times10^{-18} \mbox{ erg}\mbox{ s}^{-1} \mbox{ cm}^{-2}$, with an associated significance of $3.2\sigma$. 

\begin{figure}
    \centering
    \includegraphics[width=1.0\linewidth]{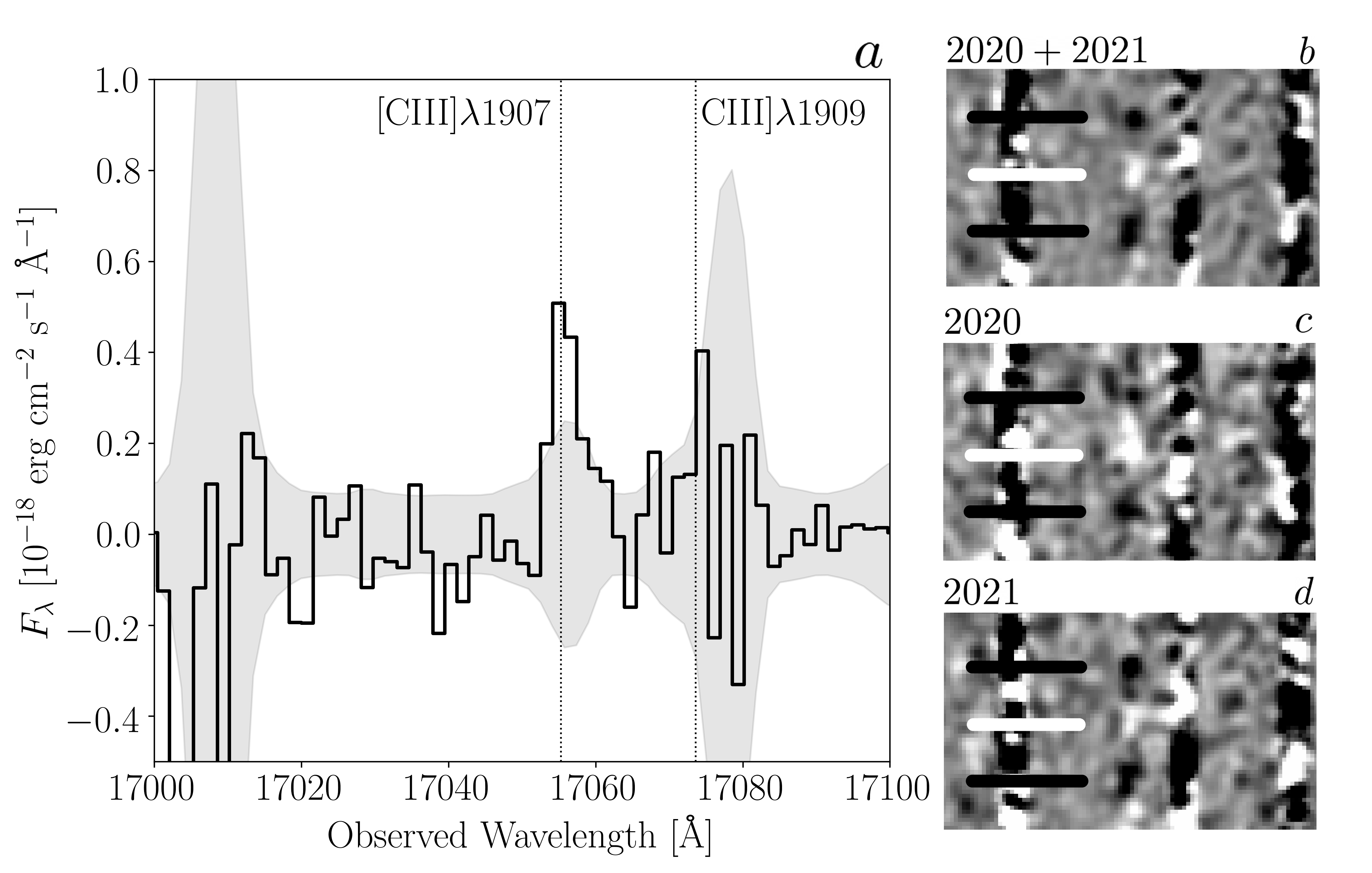}
    \caption{Reduced 1D and 2D spectra of AEGIS-33376.  (a) The 1D spectrum extracted from the \textit{all} dataset.  The $1\sigma$ error spectrum is displayed as the grey shaded region.  The vertical dotted lines show the wavelengths of [CIII]$\lambda 1907$ and CIII]$\lambda$1909 at the spectroscopic redshift of $z=7.945$. (b) Reduced 2D spectrum of AEGIS-33376 for the \textit{all} dataset showing the characteristic arrangement of a positive line surrounded on either side by a negative. The vertical positions of the positive and negative traces are indicated by white and black lines respectively. (c) Same as in panel (b) but only for data collected during 2020.  (d) Same as in panel (b) but only for data collected during 2021.  The emission line is present in the 2D spectra from both 2020 and 2021, though at lower significance than in the 2020+2021 data.}
    \label{fig:spectra}
\end{figure}

Based on the photometric redshift of AEGIS-33376, we identify this feature as one of the [CIII],CIII]$\lambda\lambda1907,1909$ doublet members. To determine if this emission line is [CIII]$\lambda 1907$ or CIII]$\lambda 1909$, we assess the expected position of the other doublet member. For an observed line of CIII]$\lambda$1909, the second doublet component would lie in a region of the spectrum clean from sky lines at $\sim17028\angstrom$.  However, given the range of anticipated [CIII]/CIII] flux ratios \citep[$1.2-1.5$;][]{Stark2017, Hutchison2019}, we would expect the line at $17028\angstrom$ to be stronger than the observed line at $17055\angstrom$, and no such emission is detected.  Therefore, the observed emission is likely associated with [CIII]$\lambda$1907, with the CIII]$\lambda$1909 doublet component expected at $\sim17073\angstrom$.  This redder component is coincident with a strong sky line, and thus we do not expect to detect it. For the observed emission of [CIII]$\lambda$1907, we establish a redshift of $z=7.945$ for AEGIS-33376, which is within the uncertainty of the photometric redshift, $z_{\rm phot}=7.8\pm0.5$. To calculate an upper limit on the flux of CIII]$\lambda1909$, we integrate the $3\sigma$ error spectrum centered on the expected location for CIII]$\lambda1909$, adopting the same width used to measure [CIII]$\lambda1907$, resulting in a 3$\sigma$ line-flux limit of  $<3.0\times10^{-18} \mbox{ erg}\mbox{ s}^{-1} \mbox{ cm}^{-2}$. Based on the $H$-band magnitude of AEGIS-33376 galaxy and the inferred continuum flux density, the [CIII]$\lambda1907$ flux and CIII]$\lambda1909$ flux limit correspond to equivalent widths of $20.3\pm6.5\rm \AA$ and $<27.3\rm \AA$, respectively. Based on the large uncertainty for CIII]$\lambda1909$ due to strong sky-line contamination, we follow \citet{Hutchison2019} in assuming a [CIII]$\lambda$1907/CIII]$\lambda 1909$ doublet ratio of 1.5, and infer a total [CIII],CIII]$\lambda\lambda1907,1909$ equivalent width of $33.7\pm 10.8\ \mbox{\AA}$. We additionally search for the [OIII]$\lambda\lambda1661,1666$ doublet, which is covered by our $H$-band spectrum based on the measured spectroscopic redshift of AEGIS-33376.  However, the derived upper limit on the [OIII]$\lambda\lambda1661,1666$ flux and the corresponding rest-frame EW limit ($<4.2\times10^{-18} \mbox{ erg}\mbox{ s}^{-1} \mbox{ cm}^{-2}$ and $<29 \mbox{\AA}$, respectively) do not place meaningful constraints on the physical properties of AEGIS-33376.
Figure~\ref{fig:spectra}(a) shows the extracted 1D spectrum of AEGIS-33376 and the $1\sigma$ error spectrum.

Due to the faint nature of the observed emission line in AEGIS-33376, and, as a result of the faint sky line that is contaminating it, we perform several tests to determine if this emission is of astrophysical origin, or if it is more likely due to sky noise or a detector artifact.  We utilize the multiple observing runs over which data was collected for this object to analyze the validity of the observed emission line at $\lambda=17055\angstrom$. First, we construct multiple datasets composed of data based on the night that the data were collected and the seeing of each individual science exposure.  These datasets are reduced using the procedure described above, and the resulting spectra are analyzed to measure the emission line at the same wavelength.  In total, we construct datasets comprising all data, the subset of data with seeing $<0\secpoint7$, data from the two nights in 2020, data from the two nights in 2021, data from 2020 with $<0\secpoint7$ seeing, and data from three nights including the 2020 data and the first night of 2021. We define the set of data containing all of the frames as the fiducial dataset, hereafter referred to as the \textit{all} dataset. The reduced 2D spectra for each of these datasets show a visible emission line at $\lambda=17055\angstrom$, which are also present in the extracted 1D spectra.  Figure~\ref{fig:spectra}(b,c,d) shows the reduced 2D spectra for the \textit{all}, 2020, and 2021 datasets.  The line is most apparent in the 2D \textit{all} spectrum, however it is visible in all three subsets. 

As an additional verification of the observed emission line, we measure the significance of the line, as well as several bright H$\alpha$ emission lines from other $z\sim1.5$ galaxies on the mask. These H$\alpha$ comparison lines typically have significances of $20-75\sigma$ when measured from the \textit{all} dataset. We perform this measurement for the same lines in each of the five constructed datasets, and test that the [CIII] line exhibits the same relative significance across the different datasets compared to H$\alpha$ from the other targets.  As the significance of very bright lines on the mask is well defined, these bright features allow for a comparison that ensures the observed emission line in AEGIS-33376 has the expected significance among the datasets, indicating it is not due to an artifact or the result of poor sky subtraction. The emission line in AEGIS-33376 follows the same trend as the bright lines, suggesting that the line is a real feature and not an artifact in the data. 

We do not detect any emission lines in the spectrum of AEGIS-3434. The spectrum of this object extends up to $\lambda = 17600\rm \AA$, which allows observation of CIII] to a maximum redshift of $z=8.22$. This  range covers the majority of redshift space expected for this object based on the photometric redshift confidence interval of $z_{\rm phot}=7.9^{+0.7}_{-0.6}$.  
However, there is some probability that AEGIS-3434 is at too high a redshift for CIII] to fall within the $H$ band.  In order to estimate the line-flux upper limit for AEGIS-3434 (and AEGIS-22108 below), which lacks a spectroscopic redshift, we use a more conservative 5$\sigma$ noise level. Within the wavelength range expected for [CIII],CIII]$\lambda\lambda1907,1909$ within the $1\sigma$ photometric redshift, half of the spectrum within this range has a $5\sigma$ limiting flux of $<2.7\times10^{-18}\mbox{ erg}\mbox{ s}^{-1} \mbox{ cm}^{-2}$ for an individual component of [CIII],CIII]$\lambda\lambda1907,1909$.  If the line is coincident with a strong sky line the flux limit can be much higher, however the $5\sigma$ limiting flux in a clean part of the spectrum is $<1.6\times10^{-18}\mbox{ erg}\mbox{ s}^{-1} \mbox{ cm}^{-2}$.  This corresponds to an upper limit on the EW of CIII] of $18.7\angstrom$ for an individual line, and $11.0\angstrom$ within clear parts of the spectrum, between sky lines.

Finally, we do not detect any emission lines for AEGIS-22108.  The placement the AEGIS-22108 slit on the MOSFIRE mask resulted in coverage to redder wavelengths, so we could observe CIII] up to a redshift of $z=8.5$. This coverage allows detection of CIII] for the full photometric-redshift confidence interval of $z_{\rm phot}=8.0^{+0.5}_{-0.7}$. Within this photometric redshift range, half of the spectrum has a $5\sigma$ flux limit of $<3.0\times10^{-18}\mbox{ erg}\mbox{ s}^{-1} \mbox{ cm}^{-2}$, corresponding to an EW limit of $<22.2\angstrom$.  Regions of the spectrum clear of sky lines for this object have a flux limit of $<1.7\times10^{-18}\mbox{ erg}\mbox{ s}^{-1} \mbox{ cm}^{-2}$, associated with an EW limit of $12.5\angstrom$. Table~\ref{tab:properties} summarizes the measurements presented here for the three $z>7$ objects in our sample.

\begin{table*}
\begin{center}
\renewcommand{\arraystretch}{1.4}
\begin{threeparttable}
\begin{tabular}{cccccccccc}
\toprule
    Field & ID    & RA & DEC &  $z_{\rm phot}$ & $z_{\rm spec}$ & Line & Line Flux & EW & $H_{\rm AB}$  \\
          & 3D-HST v4.1 &     (J2000)           &       (J2000)         &  & & & $[10^{-18}\rm \ erg \ s^{-1} \ cm^{-2}]$ & [\angstrom]   &  \\
\midrule
AEGIS & 33376 & 14:19:10.540 & +52:50:29.74 & $7.8^{+0.5}_{-0.5}$ & $7.945\pm 0.001$ & [C\textsc{iii}]1906.68 & $2.24\pm0.71$ &   $20.3\pm 6.5$ & 26.21  \\
 &  &  &  &  &  & C\textsc{iii}]1908.73 & $<3.0 (3\sigma)^{\rm a}$ &   $<27.3$ &   \\
 &  &  &  &  &  & $\rm [C\textsc{iii}],C\textsc{iii}]$ & $<5.2 (3\sigma)^{\rm b}$ &   $<47.3^{\rm c}$ &   \\

AEGIS & 3434 & 14:20:08.888 & +52:53:31.06 & $7.9^{+0.7}_{-0.6}$ & - & [C\textsc{iii}]1906.68 &  $<2.7\ (5\sigma)^{\rm a}$ & $<18.7$& 25.92       \\
&  &  &  &  &  & C\textsc{iii}]1908.73 &  $<2.7\ (5\sigma)^{\rm a}$ & $<18.7$&        \\
 &  &  &  &  &  & $\rm [C\textsc{iii}],C\textsc{iii}]$ &  $<4.6\ (5\sigma)^{\rm b}$ & $<31.8$&        \\

AEGIS & 22108 & 14:19:59.755 &  +52:56:31.44 & $8.0^{+0.5}_{-0.7}$ & - & [C\textsc{iii}]1906.68 & $<3.0\ (5\sigma)^{\rm a}$ & $<22.2$& 25.98    \\
 &  & &  &  &  & C\textsc{iii}]1908.73 & $<3.0\ (5\sigma)^{\rm a}$ & $<22.2$&     \\
 &  &  &  &  &  & $\rm [C\textsc{iii}],C\textsc{iii}]$ & $<5.1\ (5\sigma)^{\rm b}$ & $<37.7$&     \\

 \bottomrule
 \end{tabular}
 \begin{tablenotes}\footnotesize
\item[a] Upper limit estimated for an individual emission line. We adopt a 3$\sigma$ limit for AEGIS-33376, which is spectroscopically confirmed, and more conservative, 5$\sigma$ limits for AEGIS-3434 and AEGIS-22108, which only have photometric redshifts. 
\item[b] Upper limit estimated for both doublet members with fixed separation.
\item[c] Here we quote the total CIII] EW limit based on summing the detection and limit for the individual doublet members in AEGIS-33376. Elsewhere, we assume a fixed [CIII]$\lambda$1907/CIII]$\lambda 1909$ doublet ratio to infer the total EW.
\end{tablenotes}
\end{threeparttable}
 \end{center}
 \caption{Summary of galaxy properties and [CIII],CIII]$\lambda\lambda1907,1909$ measurements.}
 
\label{tab:properties}
\end{table*}

%
%
\section{Discussion and Summary}
\label{sec:disc}

We investigate doubly-ionized carbon emission in three galaxies at $z\sim8$.  While detections of CIII] are common at low redshift, they are quite rare at $z>6$ \citep{Zitrin2015, Stark2015-c3, Mainali2017, Stark2017, Laporte2017, Hutchison2019}.  We present the detection of [CIII]$\lambda$1907 in one galaxy at $z=7.945$, and upper limits for two galaxies at $z\sim8$.  Spectroscopic measurements such as these provide precise redshift measurements, and crucial constraints on the ionization mechanism and properties of the ISM during this epoch.

\begin{figure}
    \centering
    \includegraphics[width=1.0\linewidth]{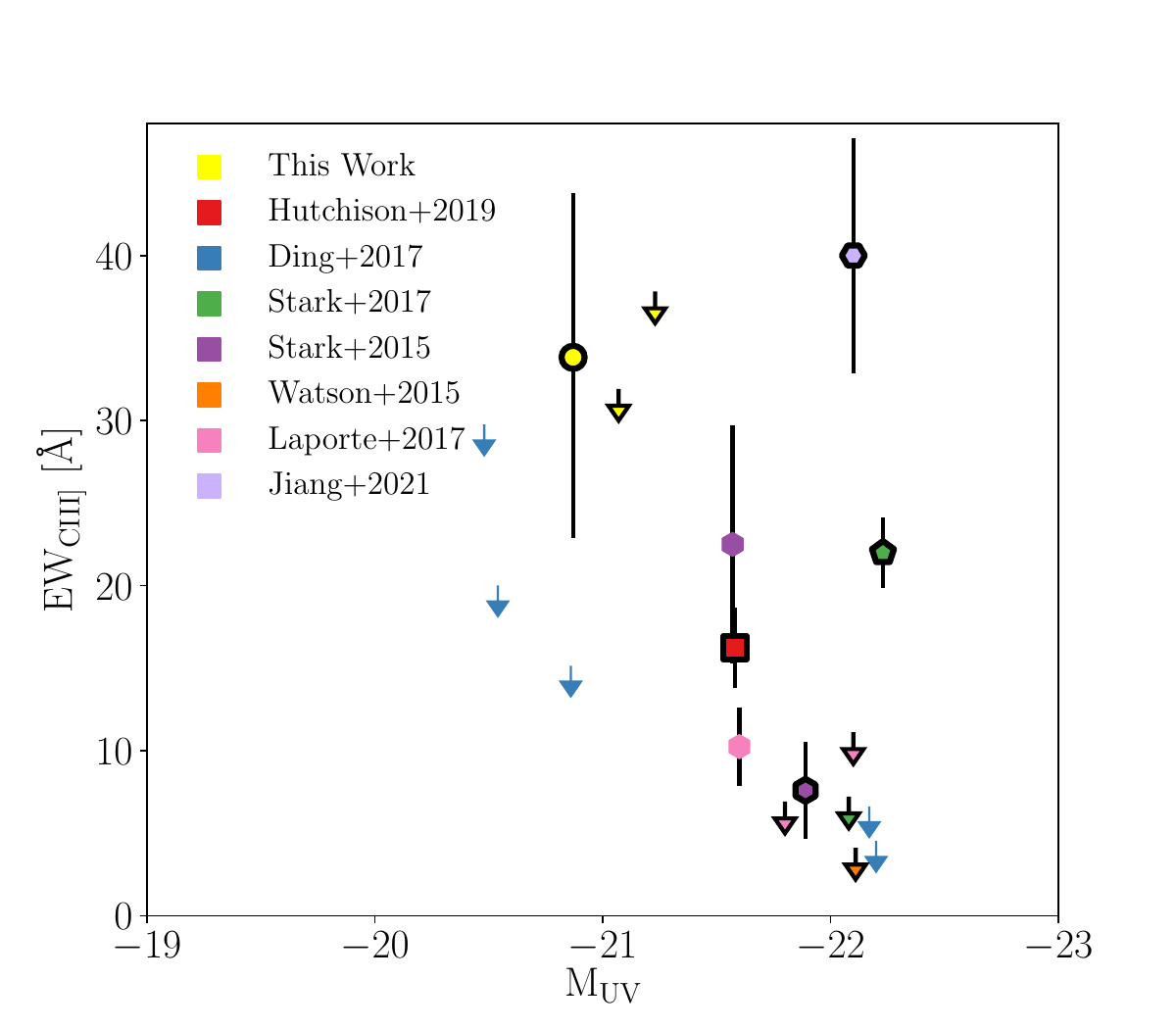}
    \caption{Comparison of total CIII] equivalent widths as a function of $\rm M_{\rm UV}$ for galaxies at $z>5$.  Objects at $z>7$ have been outlined with a black border.  The yellow markers are the three objects presented in this work. Comparison objects are shown from:  \citet{Hutchison2019, Ding2017, Stark2017, Stark2015-c3, Watson2015}, \citet{Laporte2017}, and \citet{Jiang2021}. }
    \label{fig:context}
\end{figure}

Figure~\ref{fig:context} shows the total [CIII],CIII]$\lambda\lambda1907,1909$ EW as a function of $\rm M_{\rm UV}$ for the measurements from our sample compared to other sources in the literature at similar redshifts \citep[i.e., $z>5$;][]{Watson2015, Stark2015-c3, Stark2017, Ding2017, Hutchison2019, Jiang2021}. In cases where only one of the CIII] components is detected, we assume an intrinsic [CIII]/CIII] line ratio of $1.5$, based on assuming a reasonable value for the electron density \citep{Hutchison2019}. This figure also highlights measurements of objects at $z>7$ (black outlined points).  The two upper limits presented in our sample allow for equivalent widths covering the full range spanned by strong detections at these redshifts.  The [CIII] detected for AEGIS-33376, which has the faintest $\rm M_{UV}$ of the objects in our comparison sample, represents the second largest EW, however with large uncertainties.

As CIII] is often the second strongest nebular emission line in the rest-UV \citep{Shapley2003, Mainali2017}, it provides a promising gauge of the internal physics within galaxies at high redshift.  It is also useful for measuring spectroscopic redshifts when strong-rest optical emission lines are not available at redshifts where Ly$\alpha$ becomes increasingly attenuated by the IGM \citep[e.g.,][]{Stark2010, Vanzella2011, Treu2012, Whitler2020}.  Measurements of strong CIII] observed in lower redshift samples indicate that large CIII] EWs are associated with low metallicities, high ionization parameters, and hard ionizing spectra within the ISM \citep{Erb2010, Steidel2016, Berg2016, Senchyna2017, Du2020, Tang2021}.  \citet{Stark2017} established using photoionization modelling that the large CIII] EW measured for their $z=7.73$ galaxy required models with low metallicities of $\sim 0.1 Z_{\odot}$.  In addition, utilizing measurements of Ly$\alpha$ and CIII], inferred [OIII] EW, and constraints on SiIII], \citet{Hutchison2019} demonstrated sub-solar metallicity, high ionization parameter, and the presence of young stars in a galaxy at $z\sim7.5$. The extreme CIII] observed in AEGIS-33376 is one of the largest detected EWs among these comparisons at high redshift, and it suggests similar conditions of an intense ionizing radiation field and low gas-phase metallicity. However, the ionization mechanism that is powering such large equivalent widths could also be partially explained by non-stellar processes \citep{Laporte2017, Mainali2018, Nakajima2018}.  With the redshift of this object measured, followup observations targeting Ly$\alpha$ in the $Y$ band and fainter emission lines such as SiIII]$\lambda\lambda1882,1892$ \citep{DBerg2016} could further our understanding of the conditions within this galaxy.

The small number of measured [CIII],CIII]$\lambda\lambda1907,1909$ equivalent widths observed in galaxies at $z\sim7$ appear very different from those observed in galaxies with similar $\rm M_{UV}$ at
$z\sim1-3$, which have typical EWs of $\sim 2 \angstrom$ \citep{Shapley2003, Steidel2016}. While EWs as large as $\sim20\angstrom$ are quite rare \citep[e.g.,][]{Mainali2020} at these lower redshifts, they comprise a considerable fraction of detections at $z>7$. These measurements at high redshift represent a significant evolution relative to galaxies at lower redshifts.  We expect to be most sensitive to more extreme emitters at the highest redshifts, however the low metallicities, young stellar populations, and high densities at $z\sim7$ are conducive to increased [CIII],CIII]$\lambda\lambda1907,1909$ EWs.  At yet higher redshifts we could be probing even younger and more metal-poor galaxies with higher densities.  AEGIS-33376, and the galaxy GN-z11 from \citet{Jiang2021}, are the two highest-redshift objects in our comparison, and also have the largest [CIII],CIII]$\lambda\lambda1907,1909$ EWs.  These sources may be examples of increasingly extreme star-forming conditions at increasing redshifts, though AGN contributions are also possible \citep{Laporte2017, Mainali2018, Nakajima2018}.  However, the extremely limited sample sizes preclude us from making definitive statements about the nature of galaxies at these epochs.

With current 10-meter-class ground-based telescopes and near-infrared spectrographs, the measurement of faint rest-UV metal emission lines has been limited to the most extreme line emitters. As a result, the sample sizes of such rare objects are extremely small (but can be increased with wide-field
facilities such as {\it Roman}).  Observations with {\it JWST} will probe to fainter line fluxes, allowing for measurements of more lines in the rest-UV. The redder wavelength coverage of {\it JWST} will also open up spectroscopic studies of emission lines at $z>7$ to the rest-optical.  Statistical samples probing the massive stars and ionized ISM of typical galaxies at these redshifts will provide the information necessary to understand early galaxy evolution, and the processes governing the reionization of the universe.

\section*{Acknowledgements}
We acknowledge a NASA contract supporting the “WFIRST Extragalactic Potential Observations (EXPO) Science Investigation Team” (15-WFIRST15-0004; NNG16PJ25C), administered by GSFC. We also acknowledge support from NASA Keck PI Data Awards, administered by the NASA Exoplanet Science Institute. RE acknowledges funding from JWST/NIRCam contract to the University of Arizona, NAS5-02015. Data presented herein were obtained at the W. M. Keck Observatory from telescope time allocated to the National Aeronautics and Space Administration through the agency's scientific partnership with the California Institute of Technology and the University of California. The Observatory was made possible by the generous financial support of the W. M. Keck Foundation. 
We wish to extend special thanks to those of Hawaiian ancestry on whose sacred mountain we are privileged to be guests. Without their generous hospitality, most of the observations presented herein would not have been possible.

\bibliographystyle{apj}
\bibliography{ciii_z8_letter}

\end{document}